\documentclass[twocolumn,showpacs,prc,nofootinbib]{revtex4-1}

\def\be{\begin{eqnarray}}
\def\ee{\end{eqnarray}}

\def\roughly#1{\mathrel{\raise.3ex\hbox{$#1$\kern-.75em%
\lower1ex\hbox{$\sim$}}}}

\begin{document}

\title{Back-to-back pair correlation of  Majorana neutrinos with transit magnetic moments}

\author{Hyun Kyu Lee}
\affiliation{%
Department of Physics, Hanyang University, Seoul 133-791, Korea \& \\
Asia Pacific Center for Theoretical Physics, Pohang 790-784, Korea
}

\date{\today}

\begin{abstract}
     The pair production of Majorana neutrinos with transit magnetic moments  from the annihilation of charged particles in colliding experiments  is discussed using the Pauli interaction, through which the neutral neutrinos but with magnetic moments can be probed by photon.  The  pair of neutrinos with different flavors  are produced due to the transit magnetic moment coupling. We discuss the correlations of flavors in pairs produced back-to-back in the center of mass frame, where the angular distribution peaks at $\theta=\pi/2$ with respect to the beam direction.  We demonstrate that the flavor mixing angle can be inferred by measuring the flavor correlation in pairs.

\end{abstract}

\pacs{13.40 Em, 13.66 Hk, 14.60 St}

\maketitle

Among the fundamental building blocks of Universe in the standard model, neutrinos are electrically neutral particles with spin 1/2, which are interacting only weakly being classified as upper component of weak left-handed   doublets.
The atmospheric and  solar neutrino observations\cite{ashie} \cite{aharmim}  as well
as reactor experiments\cite{Araki} show  the evidences of oscillations between
different flavors of neutrinos, which are only  possible if neutrinos are  massive and  flavors are  mixed in the mass eigenstates.
Neutrinos do not have electric charges  but they are found to have non vanishing mass and  it is natural to ask the possibility of magnetic moments\cite{giunti} through which they can interact with photons directly.    We assume, in this work,  the case where the massive neutrinos have non vanishing magnetic moments or the transition magnetic moments, which  can induces a spin-dependent coupling to photons.

The bounds for the neutrino magnetic moments  obtained from the
experiments\cite{wong} \cite{gg}\cite{elmfors}\cite{ag} and theoretical considerations\cite{bell} are varying in wide range of  $10^{-15} - 10^{-7} \mu_B$,  where  $\mu_B$ is the Bohr magneton.
In the standard model, the neutrino magnetic moment induced by one-loop effect \cite{fujikawa}
is $
\mu_{\nu} = 3 \times
10^{-19} \left(\frac{m_{\nu}}{eV}\right)\mu_{B}$,  which is
much smaller than the above bounds. It is also worth mentioning that the upper bound for the magnetic moments are less stringent for Majorana neutrinos than for Dirac neutrinos.

In the lowest order of standard model, the neutrino pair production by  the annihilation of charged particles is through $Z^0$ channel. However if neutrinos have non-vanishing magnetic moments\footnote{ It is interesting to  note that  the effect of magnetic moments of neutrinos on the
vacuum instability  has been recently investigated in the presence of strong
external magnetic field to  find out   that with non vanishing magnetic moment there appears vacuum instability beyond
critical field strength, $B_c = \frac{m_{\nu}}{\mu_{\nu}}$  against the pair production of
neutrinos\cite{yoon}\cite{hklyoon}} they can be also produced through photon channel as well\cite{barut}\cite{schgal}\cite{deshpande}.  When we adopt the Pauli interaction\cite{pauli} as an effective interaction as for beyond-standard model physics,
the cross section of pair production becomes dominated by the Pauli interaction over the standard process through $Z^0$ channel as the energy is increasing. For example,  if magnetic moments are not much smaller than $10^{-10} - 10^{-9} \mu_B$, substantial increases of pair production rates at Large Hadronic Collider(LHC, $E^{CM}_{LHC} > 10 TeV$)  and Ultra High Energy Cosmic Ray experiments(UHECR, $E^{CM}_{GZK} \sim 100 TeV$ ) are expected\cite{GLPY}.

 Majorana neutrinos are known to have only transit magnetic moments, which implies that the lepton flavor numbers are not conserved  in this process.
Therefore  if neutrinos produced are  Majorana  type, then the pairs
 should be produced with different flavors.  It  gives us an interesting possibility  to  figure out  which type of     neutrinos is involved,   Majorana or Dirac.
In this work, we discuss the flavor correaltions in a pair of neutrinos, which can be used to infer the flavor mixing angles for  Majorana neutrinos.  Since the angular distribution of produced pairs through magnetic moment coupling peaks at $\theta =\pi/2$ with respect to the beam direction, the events observed  at the right angle in  the center of mass frame can be easily distinguished from those of  standard model process.

The Majorana field is basically represented by two-component spinor, $\chi$. For a free particle
the Lagrangian of two component Majorana field is given by
\be
{\cal L}= \chi^{\dagger}\bar{\sigma} \cdot \partial \chi -\frac{m}{2}\left[\left(\chi^{C}\right)^{\dagger}\chi+\chi^{\dagger}\chi^{C}\right].
\ee
where
\begin{eqnarray}
i \bar{\sigma} \cdot \partial \chi - i m \sigma^{2} \chi^{\ast} = 0. \label{Majoranaeq}
\end{eqnarray}

 Using the four component Majorana field, $\Psi(x)$,  the interaction Lagrangian for the Majorana neutrino with  Pauli interaction can be written down,
\begin{eqnarray}
{\cal L}_{int} &=& i \frac{\mu_{ij}}{2} \bar{\Psi}^{i}_{M} \sigma_{\mu\nu} \Psi^{j}_{M} F^{\mu \nu} ,  \label{LagM}
\end{eqnarray}
where $\sigma_{\mu\nu}=\frac{i}{2}\left[\gamma_{\mu},\gamma_{\nu}\right]$ and $g_{\mu\nu} = \left(+,-,-,-\right)$.  $\mu_{ij}$ is a transition magnetic moment which is anti symmetric for a Majorana neutrino, $\mu_{ij}=-\mu_{ji}$.

The  process we are considering is  the  annihilation of charged
fermion   into the neutrino pair through photon channel with  Pauli interaction at very high energy. In the high energy region, where  the particle masses are very small compared
 to the energy scale, $m_i \ll E$,  the differential cross section and total cross section respectively converge to simple expressions\cite{GLPY}:
\begin{eqnarray}
\left(\frac{d\sigma}{d\Omega}\right)&=&\frac{\alpha Q^2\mu^2_{12}}{4\pi} \mathrm{ sin^2 \, \theta}\label{e53}
\end{eqnarray}
and
\begin{eqnarray}
\sigma = \frac{2\alpha Q^2 \mu^2_{12}}{3},\label{e54}
\end{eqnarray}
which are similar to the results for a Dirac neutrino with magnetic moment\cite{barut}.
One can see that the angular distribution peaks at $\theta=1/2$.  It is compared to
the angular distribution of standard model process, which  has  maximum for  $\theta = 0$ and $\pi$  and minimum
for $\theta  \sim  \pi/2$.  These features are quite different from those  with Pauli interaction.  To get some idea on the energy scale for which the magnetic moment coupling is dominant,  we can define  the energy scale, $E_{0.1}$, for which the total cross section  becomes $\sim 10$\% of standard model,
\begin{eqnarray}
E_{0.1} \sim 10^2 \left(\frac{10^{-10}\mu_B}{\mu}\right) TeV.
\end{eqnarray}

Then the detectors located around  the right angle to the beam direction can measure the back-to-back correlations in pair production, where the production rate is supposed to be maximum.  Most of detectors are using electromagnetic triggers at the end stations, which implies electrons or muons are those to be detected finally.  The Majorana pairs in   mass eigenstates, $\nu_1$ and $\nu_2$, are produced and interact weakly with other particles to produce charged leptons to be detected.  We consider only two mass eigenstates to simplify the situation.  $\nu_1$ and $\nu_2$ are linear combinations of weak eigenstates,    $\nu_{\alpha}$ and $\nu_{\beta}$,
\be
\nu_1 &=& \cos \delta~ \nu_{\alpha} + \sin \delta ~\nu_{\beta} \nonumber \\
 \nu_2 &=& -\sin \delta~ \nu_{\alpha} + \cos \delta ~\nu_{\beta}, \label{mixing}
\ee
where $\delta$ is a mixing angle and $\alpha$ and $\beta$, for example, can be electron and muon.
Now consider two targets A and B placed at the opposite side of the center of mass at the right angle to the beam direction. Suppose $\nu_1$ hits the target A, then  the rates, $N_A^{(1)}(\alpha)$ and $N_A^{(1)}(\beta)$ for  detecting of  $l_{\alpha}$ and $l_{\beta}$ respectively,   at the detectors surrounding A are proportional to the mixing angles,
\be
N_A^{(1)}(\alpha) =N \cos^2\delta, ~~ N_A^{(1)}(\beta) =N  \sin^2\delta, \label{na1}
\ee
where $N$ is introduced as an overall normalizing factor.
The second neutrino, $\nu_2$, produced in the opposite direction hits the target B to produce leptons,
\be
N_B^{(2)}(\alpha) =N \sin^2\delta, ~~ N_B^{(2)}(\beta) =N  \cos^2\delta. \label{nb2}
\ee
On the other hand when  $\nu_2$ hits the target A we get
\be
N_A^{(2)}(\alpha) =N \sin^2\delta, ~~ N_A^{(2)}(\beta) =N  \cos^2\delta, \label{na2}
\ee
and
\be
N_B^{(1)}(\alpha) =N \cos^2\delta, ~~ N_B^{(1)}(\beta) =N  \sin^2\delta. \label{nb1}
\ee

Since the rate of producing  $\nu_1$ in the direction A is the same as for $\nu_2$ simply because they are simultaneous events, the rates of producing  $l_{\alpha}$ or $l_{\beta}$  at each targets are not depending on the mixing angle,
\be
N_A(\alpha) &=& N^{(1)}_A(\alpha) + N^{(2)}_A(\alpha) = N \nonumber \\
N_B(\alpha) &=& N^{(1)}_B(\alpha) + N^{(2)}_B(\alpha) = N \label{nab}
\ee
and the same for $l_{\beta}$.

However the back-to-back correlation, $R$, defined by the product of the rates of $l_{\alpha}$ at A and  $l_{\beta}$  at B,
\be
R &\equiv& [N_A^{(1)}(\alpha) \times N_B^{(2)}(\beta) +  N_A^{(2)}(\alpha) \times N_A^{(1)}(\beta)]/N \nonumber \\
&=& \cos^4\delta + \sin^4\delta = \frac{1}{2}( 1 + \cos^2 2\delta),\label{R}
\ee
turns out to be strongly dependent on the mixing angle, $\delta$, to be measured.
The maximum value, $R=1$,  is obtained  for the cases of no-mixing, $\delta = \frac{n\pi}{2}(n = 0,1,2, \ldots)$.  But interestingly the minimum value, $R= 1/2$, is obtained when the mixing angle is $\pi/4$ for the maximal mixing. And for moderate mixing, $\delta =\pi/8$, one can get $R=3/4$.

In summary, we discuss an observational possibility of mixing angle of Majorana neutrinos by measuring back-to-back correlation of flavors in the pairs produced assuming that the Majorana  neutrinos  have non-vanishing transit magnetic moments and they are produced in pairs from the  annihilation of charged particles through Pauli interaction.  It turns out that  because of the periodicity of the correlation $R$  with a period of $\pi/2$ in mixing angle it can measure  $\delta$  effectively modulo  between  $0$ and $\pi/4$ for which $R$ ranges from 1 to 1/2, which can be of interest particularly for the moderate mixing angles up to maximal mixing.  Although the present energy scale might  not be sufficiently high enough to be effective in measuring the correlations for the  magnetic moment smaller than $\sim 10^{-10} \mu_B$ and moreover most of the present experimental systems are not so sensitive enough for the neutrino detections, we suggest that the neutrino magnetic moment with Pauli coupling  can open an interesting channel in the future experiments at higher energy with more sensitive detecting systems of neutrinos.   Finally it should be noted that the Pauli coupling  is a kind of effective interaction term which might be valid only up to some scale, which  we assume to be higher than the scale we are considering in this work.

\subsection*{Acknowledgments}
The author would like to thank B.-G. Cheon,  Y. Goh, W-G. Paeng and Y. Yoon  for useful discussions.   This work is supported by the
 WCU project of Korean Ministry of Education, Science and Technology (R33-2008-000-10087-0).

\end{document}